\documentclass[manuscript,12pt]{rsauthor}
\usepackage{graphicx}
\usepackage{epstopdf}
\usepackage{fullpage}
\usepackage{natbib}
\usepackage{amsmath}
\usepackage{lineno}
\usepackage{longtable}
\usepackage{multirow}
\usepackage{lscape}
\usepackage{rotating}
\usepackage{url}
\usepackage[T1]{fontenc}
\usepackage{mathptmx}

\usepackage{color}

\begin{document}

\thispagestyle{empty}

\Large
\begin{center}

\bigskip

\textbf{Local properties of patterned vegetation:
quantifying endogenous and exogenous effects}

\bigskip

\normalsize
Gopal G. Penny $^{1,*}$,
Karen E. Daniels$^2$,
Sally E. Thompson$^{1}$ \end{center}

\normalsize
\noindent [1] Department of Civil and Environmental Engineering, University of
California, Berkeley, California, 94710, sally.thompson@berkeley.edu\\
\noindent [2] Department of Physics, North Carolina State University, Raleigh,
NC 27695, kdaniel@ncsu.edu \\

\noindent \textbf{*Corresponding author:}
Gopal G. Penny\\
Department of Civil and Environmental Engineering, \\
University of California, \\
Berkeley, California, 94710\\
Email: gopal@berkeley.edu\\
Phone: (+1) 510 642 1980\\
Fax: (+1) 510 643 5264

\bigskip

\noindent \textbf{Running title:} Local properties of patterned vegetation


\noindent \textbf{Manuscript type:} submitted to Special Issue on ``Pattern
Formation in the Geosciences''

\vspace{1cm}

\noindent \textbf{Abstract}\\

Dryland ecosystems commonly exhibit periodic bands of vegetation,
thought to form due to competition between individual plants for
heterogeneously distributed water.  In this paper, we
develop a Fourier method for locally identifying the pattern
wavenumber and orientation, and apply it to aerial images from a region of
vegetation patterning near Fort Stockton, Texas.
We find that the local pattern wavelength and orientation are typically
coherent,
but exhibit both rapid and gradual 
variation driven by changes in hillslope gradient and orientation, the potential
for water accumulation, or soil type.  Endogenous pattern dynamics, when
simulated for 
spatially homogeneous topographic and vegetation conditions, predict pattern
properties that are much less variable than the orientation and wavelength
observed in natural systems. Our local pattern analysis, combined with
ancillary datasets describing soil and topographic variation, highlights a
largely unexplored
correlation between soil depth, pattern coherence, vegetation cover and pattern
wavelength.  It also, surprisingly, suggests that downslope
accumulation of water
may play a role in changing vegetation pattern properties.

\newpage
\linenumbers
\modulolinenumbers[5]

\section{Introduction \label{sec:intro}}  

Pattern formation occurs in numerous ecological and biological systems, where it has been linked to reaction-diffusion (Turing) type instabilities \citep{Rietkerk-2002}, hydrodynamic instabilities \citep{Thompson-2010grass}, and potential (variational) dynamics that maximize or at least increase ecological productivity \citep{Lefever-2009, Pringle-2010}.  Patterns have been observed in dryland vegetation \citep{Borgogno-2009}, in bogs and wetlands \citep{Rietkerk-2004Bog, Larsen-2011}, in mussel beds \citep{vandeKoppel-2008}, termite mounds \citep{Pringle-2010} and other systems \citep{Rietkerk-2004}.  In many cases ecological patterns form an intermediate realization between environmental states in which the entire landscape is either colonized or bare.   As such they often indicate the presence of bistable states, which are characterized by the potential for critical and locally irreversible transitions \citep{Scheffer-2009}.  Observations and theoretical work both suggest that transitions between vegetated and desertified (unvegetated) conditions in patterned systems are preceded by striking changes in the morphology of the vegetation patterns, which thus act as an early-warning sign of deteriorating ecosystem health \citep{Kefi-2007}.  Understanding the controls on the morphology of vegetation patterns therefore has practical interest in terms of ensuring that observed changes are interpreted correctly.

Vegetation pattern formation in dryland ecosystems is a  global phenomenon, ranging
from random distributions of bare soil and vegetation canopies
\citep{Caylor-2004} to highly organized spatial distributions with identifiable
length-scales and orientations \citep{HilleRisLambers-2001, Rietkerk-2002,
Borgogno-2009}. Intermediate cases such as power-law (scale-free) clustering
\citep{Scholes-1997, Scanlon-2007, VonHardenberg-2010}, and dendritic
structures in which vegetation concentrates along drainage lines
\citep{McGrath-2012, Thompson-2011a} are also observed.
All of these morphologies can be related to the presence, strength and
directionality of positive feedbacks 
that  concentrate resources (such as nutrients, soil carbon and
water) that sustain plant life in a localized region near the plants
\citep{Ravi-2008, Greene-1992, Galle-1999, Harman-2012, Puigdefabregas-2005, Schlesinger-1990}.
These feedbacks have lead to the moniker `ecosystem engineers' being
applied to perennial plants in dryland ecosystems: they create the
conditions necessary for their own survival \citep{Gilad-2004,Yizhaq-2005}.

The formation of periodic vegetation patterns is strongly linked to the coupling of
(i) water redistribution to the vicinity of plants, and (ii) competition between individual plants for
access to this water resource \citep{Borgogno-2009, Bromley-1997, Lefever-2009, Seghieri-1997}. The dynamics of patterned
vegetation systems has formed a focus of ecohydrological and nonlinear dynamics
research, motivated by the increasingly well-demonstrated connection between
pattern morphology and desertification risk \citep{Rietkerk-2002, Kefi-2007, Barbier-2008,
Deblauwe-2011}; and by the inherently interesting nonlinear dynamics of these
systems \citep{Lefever-2009, Meron-2004}.  Field and theoretical work has identified the important roles of vegetation dynamics
\citep{Kefi-2007}, seed dispersal processes \citep{Thompson-2008, Thompson-2009}, root morphology \citep{Barbier-2008, Lefever-2009}, surface flow dynamics \citep{Thompson-2011b} and climate feedbacks \citep{Konings-2011} in modulating pattern morphology.

While our understanding of vegetation pattern dynamics is improving, there remain
undeniable differences between simulated vegetation patterns and the natural
ones observed in the field. Natural patterns are characterized by a degree of
disorder and heterogeneity on multiple scales that is not reproduced
in idealized models. Disorder can arise intrinsically through the
existence of a range of stable wavelengths and orientations, with transitions in
space and time occurring through local pattern pattern defects \citep{Cross1993}.
In addition, the pattern wavelength and orientation can be strongly influenced
by either wavelength-scale inhomogeneities \citep{Lowe1983} or larger gradients
\citep{Eckhaus1997} in the underlying system, as well as the presence of
boundaries \citep{Hu1993} or noise \citep{Lindner2004}.

In natural systems, this means that disorder observed in vegetation patterns
could either reflect
intrinsic features of the pattern forming process ({\itshape endogenous
effects}), or could reflect spatial
changes in soil structure and local topography  ({\itshape exogenous
effects}). For instance,
\citet{Thompson-2008} explored whether the unrealistically smooth nature of many
models of vegetation biomass distribution was an artifact of representing seed dispersal as a diffusive process. Greater fidelity between
modeled and observed (disordered) vegetation patterns was achieved by
representing plant population migration with a seed dispersal kernel. More
simply, theoretical treatments of vegetation patterning usually neglect variations in soils and topography, or impose periodic boundary-conditions
that remove the differences in water availability between the top and bottom of a hillslope.
Such simplifications will inevitably lead to idealized representations of
the vegetation response and obscure interactions between the intrinsic pattern structure and the spatial structure of the landscape. For example, \citet{McGrath-2012} demonstrated that the orientation and wavelength of vegetation patterns modeled with realistic boundary conditions changed between the top and bottom of the hillslope.  In control simulations with periodic boundary conditions, vegetation bands formed with a single wavelength and were orientated at 90$^\circ$ to the direction of water flow, in agreement with other modeling studies. When realistic boundary conditions were imposed, however, the bands near the top of the hillslope curved to lie perpendicular to the ridgeline.  Similarly, the effects of spatial heterogeneity in soil properties on pattern formation have not been widely explored (although see \citet{Thompson-2008}).  A number of theoretical studies indicate that local biomass, band properties and band coherence should vary with changing minimum and maximum infiltration capacities and soil properties \citep{Ursino-2006, Thiery-1995}, and there are some tantalizing hints that subsurface features, such as the ironstone underlying tiger bush in Niger, calcrete hardpans underlying banded patterns in Texas, and silcrete hardpans underlying \textit{mulga} bands in Australia \citep{McDonald-2008, White-1970, Mabbutt-1987}, may have an association with patterned vegetation.

Predictions about the interaction of vegetation patterns with changing soil or topography can be investigated using remotely sensed datasets, an approach with a long and growing history. Vegetation patterns in Africa were first observed from light aircraft flights \citep{Macfadyen-1950, Worrall-1960}. Initial analyses of the patterns demonstrated their spatial regularity on the basis of the two-dimensional Fourier power spectrum \citep{Couteron-2001}.  More recently, \citet{Deblauwe-2011} undertook large scale analyses of morphological trends in vegetation patterns in the Sudan by linking several remotely sensed datasets: surface imagery from the
System for Earth Observation (SPOT) satellite, topographic data from the Shuttle Radar Topography  Mission (STRM), and
rainfall data from the Tropical Rainfall Measuring Mission (TRMM). This allowed
analysis of pattern morphology at 400~m resolution over multiple square kilometers.  Although higher
resolution datasets are available, they have generally either been analyzed only
over relatively small spatial scales ($\approx 1$~km$^2$) \citep{Barbier-2008} or investigated from the perspective of identifying morphological change over time \citep{Deblauwe-2012}.

A key open question is therefore to characterize small-scale irregularities in
vegetation patterns, and to classify them based on whether they arise due to
either endogenous dynamics or variation in exogenous features
imposed by the landscape. Understanding the implications of such variation on
the resilience and stability of the ecosystems has important consequences for
identifying and preventing potential desertification.

In this paper, we report on the development of methods suitable for quantifying
local patterns within high-resolution aerial photography, and for relating
those features to ancillary datasets describing soil and topographic variation.
Our site, located near Fort Stockton, Teaxs (see \citet{McDonald-2008}) was
selected because it exhibits vegetation bands and also offers several
characteristics that facilitate studies of covariation.
First, the aerial photography covers a large region at a
resolution (0.5 and 1~m) comparable to the diameter of perennial vegetation
canopies, allowing for the observation of individual plants.
Second, high resolution (10~m spatial, 15~cm vertical) elevation datasets are
available from the US National Elevation Dataset, allowing changes in the orientation and gradient of the hillslope to be mapped on scales that are much less than the pattern wavelength.
Finally, the study area has been mapped as part of the SSURGO national soils
database, and contains considerable variation in soil type.  The availability of these datasets provides
a valuable opportunity to investigate correlations
between vegetation patterning and soil characteristics over tens of square kilometers. 

Our analysis depends on these spatial datasets, and is subject to their inherent limitations.  These limitations include the resolution, spatial artifacts, and the risk of spurious correlation, given that vegetation features are often considered when mapping soil boundaries.  To minimize the effects of issues, we focus on two large-scale hypotheses:
(1)  The wavelength and orientation of the vegetation pattern are locally coherent but exhibit
both rapid and gradual variation; and
(2) The variability in vegetation pattern features will correlate with soil and elevation features in predictable ways.
We note that by focusing on soil and topographic features, we inherently assume that spatial variations in pattern morphology arise due to spatial heterogeneity in landscape properties, rather than due to the nonlinear dynamics of the pattern forming process itself.  In patterns far from threshold, features such as defects, dislocations, grain boundaries, and boundary conditions can result in spatial heterogeneity in pattern properties, \textit{even} when all other fields are homogeneous \citep{Cross-2009}.  To control for this possibility, we also briefly address the following null hypothesis:
 (3) Large-scale variations in pattern properties can be explained by the nonlinear interactions associated with vegetation pattern formation.

We test the major hypotheses through the development of a localized
Fourier technique to identify pattern wavenumber and direction.  To address
Hypothesis 1, this technique was applied to high (0.5 m) resolution  aerial
photography.  To address Hypothesis 2, the technique was applied to a 188 km$^2$ area,
allowing local pattern properties to be correlated to the site soil and topographic properties.
Hypothesis 3 was tested by identifying a subset of the 0.5 m resolution image containing banded patterns that were close to ideal (i.e. reproducible by a model).  We applied the Fourier analysis technique to both observed and simulated patterns, and compared variability in the wavenumber and direction fields in the modeled and observed patterns.

We find that our two major hypotheses are satisfied. Local patterns are oriented in
the same direction as the topographic slope and the pattern wavelength decreases
for for steeper gradients.  Deviations from these trends are associated with the presence of ridges, stream
channels, anthropogenic features or changes in soil type.  Different soil types
within the study area determine the pattern boundaries and the pattern
morphology: shallow soils are associated with highly coherent, shorter
wavelength patterns, and deep soils with patterns that become incoherent and
increase in wavelength near stream channels.  Finally, the modeling test validated our decision to focus on landscape and soil features. Even in the most uniform region of the observed patterns, the variability in the real pattern properties exceeded that which could be retained in the steady state solution of a physical model that closely reproduced the mean pattern properties.

\section{Methods \label{sec:methods}} 

\subsection{Study Site and Data}
\begin{figure}[h]
\caption{(a) Binary image of the Fort Stockton region, covering an $\approx$
188 km$^2$ area. Vegetation is shown as black and bare soil is shown as white.
(b) Detail of size  $260\times 260$ m$^2$, showing the original detailed image.
(c) Binary image detail.}
\label{fig:binary}
\includegraphics[width=\linewidth]{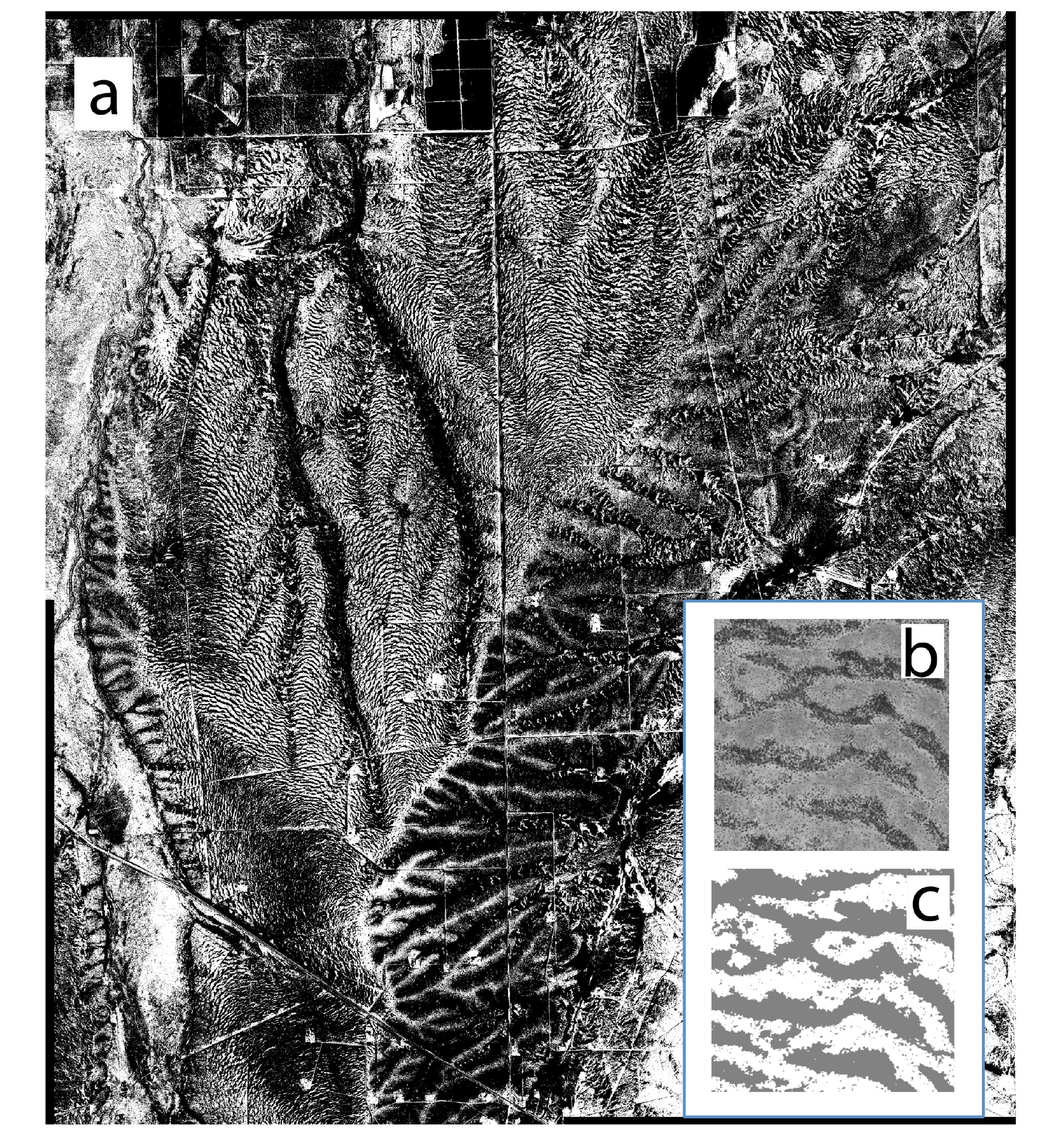}
\end{figure}

The study site is a 188~km$^2$ area located approximately 30 km NW of Fort
Stockton, Texas (coordinates: 31$^\circ$05' N, 103$^\circ$03' W)
\citep{NCDC-2010}.  The climate is hot and dry, receiving 370~mm annual rainfall
on average, mean summer maximum temperatures of approximately 35$^\circ$C and
winter minima near freezing.  The site is part of a large cattle ranch and is
used for grazing.  Dominant vegetation species include tarbush
(\textit{Flourensia cenua}), bunch and sod grasses (\textit{Aristida purpurea,
Bouteloua curtipendula} and \textit{Scleropogon brevifolius}), and mixed mesquite
(\textit{Prosopsis glandulosa}) and juniper (\textit{Juniperus pinchotti}) brush
\citep{McDonald-2008}.  The site contains a striking spatial
pattern consisting of bands of continuous vegetation cover lying over bare soil
\citep{McDonald-2008}; see Fig.~\ref{fig:binary}.

High resolution imagery (0.5~m and 1~m pixels) were obtained from Digital
Globe, and from the National Agricultural Imaging Project
\citep{NAIP-2010}.  We used the highest resolution images from Digital Globe for
fine-scale analysis and a comparison between modeled and observed pattern
morphology.  We analyzed the NAIP images, which cover the whole
area, to relate local pattern properties to soil and topographic features.

We classified the image pixels as `vegetation' or `no
vegetation' using a supervised classification based on total brightness
\citep{Richards-1999}.  We used brightness because the perennial vegetation was
not actively photosynthesizing when the photographs were taken, meaning that
standard vegetation indices could not discriminate the vegetated locations. An
example of the resulting binary image is shown in Fig.~\ref{fig:binary}. Darker colors represent vegetation cover and the
lighter colors bare soil. The insets show an original and classified image over
a 260 $\times$ 260~m$^2$ window.

\subsection{Fourier Windowing Method \label{sec:FWmethod}}

To quantitatively test our hypotheses, we developed a quasi-local technique to
measure pattern wavelength $\lambda$ and pattern orientation $\Theta$ for the
binary images.  The technique provides information about the local wavevector
$\vec{k}=k_x \hat{x}+k_y\hat{y}$, similar to that provided by short-time
Fourier transforms or wavelet based approaches used in timeseries analysis
\citep{Allen-1977, Daubechies-1990}. Local wavevectors are useful
for classifying convection patterns \citep{Heutmaker1985} and for identifying
pattern defects \citep{Egolf-1998, Daniels-2008-CBO}. Here we applied a
two-dimensional Fourier transform to obtain the power spectrum within a
square, moving window. Local wavelength and pattern orientation were identified for each window, and the
results averaged for all windows that spanned a given location. This
straightforward technique is suitable for identifying the dominant pattern
properties in noisy images with irregular patterning. More elegant and
rigorously-local techniques, based on the ratios of the spatial derivative of
the banding pattern \citep{Egolf-1998}, could not be applied because the
vegetation bands deviate substantially from a sinusoidal pattern in space. The drawbacks of the
short-time Fourier transform---namely the tendency to truncate long wavelengths
due to the finite window size, and poor localization of short-wavelength
components \citep{Pinnegar-2004}---do not pose significant difficulties
in the current
application, provided that the window size is greater than the local
wavelength of the pattern.

For each window of size  $L \times L$, we obtained the 2D fast Fourier transform
${\tilde f}(k_x, k_y)$  of the pattern $f(x,y)$. For each window's ${\tilde
f}(\vec{k})$, we calculated the power spectrum $S({\vec k})= |{\tilde f}({\vec
k})|^2$. As $L$ increases, the likelihood of identifying a single wavelength and
orientation decreases. Conversely, as $L$ decreases, the $k$-resolution in Fourier space is reduced. To optimize both the localization of
measurements and the resolution, we chose $L$ to be at least 5$\lambda$. The window was applied
to overlapping regions of the pattern at 20~m
intervals.

The power spectrum measures the power
contributed to the pattern by each wavevector ${\vec k}$. To separate the local
wavelength from its orientation, we decomposed each ${\vec k}$ into its
magnitude (wavenumber) $k = \sqrt{k_x^2+k_y^2} = 2 \pi / \lambda$ and its
orientation $\theta$.  Due to symmetry, the pattern orientation can
be defined only between 0 and $\pi$.
To identify the dominant $k$ in each window, we binned $S({\vec k})$ into
annular rings of width $k=6\pi/L$. To deconvolve the natural $1/k$ scaling of
the image \citep{Burton-1987, Tolhurst-1992}, we computed the total power within
each annular ring, $S(k)$. The location peak of this total power
(rather than the mean power) is used to define the most
energetic wavenumber, $k_1$. To compensate for the large bin width, we computed
the location of the weighted average $k_1 \equiv \langle k S(k) \rangle /
\langle S(k) \rangle$ over all rings that formed part of the peak and contained
$>75\%$ of the peak power. We discarded windows where $k_1$ corresponded to the
limits of the window function or pixel resolution, and, to avoid discarding
sites where $L$ truncated $\lambda$, visually confirmed that these windows
corresponded to un-patterned areas. Finally, we discarded windows where the mean
power of the pattern-forming wavenumber was less than that of noise. To
determine the dominant pattern orientation, we binned
$S({\vec k})$ into segments of width $\pi/8$ and computed the average power,
$S(\theta)$. The dominant orientation was located via the
weighted average $\theta_1 \equiv \langle \theta S(\theta) \rangle / \langle
S(\theta) \rangle$. The most energetic values $(k_1, \theta_1)$ for each window
were
used to generate spatial maps $\lambda(x,y)$ and $\Theta(x,y)$ representing the
local pattern wavelength and orientation. For each  $x,y$ location, the assigned value
of $\lambda$ and $\Theta$ comes from the average $2 \pi/k$ and $\theta$ of all
windows which include that $x,y$.

Although the procedure so far assumes a two-dimensional pattern with a single wavelength
and orientation within each window, the power spectra regularly contained
additional peaks.  We evaluated the  uniqueness of the local pattern properties
by comparing the dominant peak to its most distant energetic peak.  Energetic
peaks were defined as those containing
$>75\%$ of the power of the dominant peak.  The most distant peak was then
defined as the energetic peak with wavenumber $k_2$ and orientation $\theta_2$
that maximized $|k_1 - k_2|$ and $|\theta_1 - \theta_2|$.  Uniqueness  metrics
were then defined for both the wavenumber and the orientation as:
\begin{align}\label{eq:Quality}
	Q_k & = 1 - \frac{|k_1-k_2|}{\mathrm{max}\left(k_1,k_2\right)}  \\
\nonumber 	Q_\theta &= 1 - \frac{|\theta_1-\theta_2|}{\pi/2}
\end{align}
\noindent $Q_k$ and $Q_\Theta$ quantify the degree to which the dominant peak is
either a unique energetic peak
($k_1 = k_2$ and $\theta_1 = \theta_2$ implies $Q_k=1$ and $Q_\theta=1$)
or is one of at least 2 energetic peaks lying orthogonal to
each other ($Q_\theta = 0$) or separated by a large relative distance in $k$ space.

Matlab code that performs the Fourier Windowing Method outlined in this section is provided as online supplementary material.

\subsection{Construction of the spatial fields used for analysis}

\begin{figure}[h]
\caption{
Maps of (a) local wavelength $\lambda(x,y)$, with color scale indicating
wavelength in meters and (b) local orientation $\Theta(x,y)$, with color scale
indicating orientation in radians. Areas in white do not contain patterns as
recognized by the Fourier windowing method.}
\label{fig:fourier}
\includegraphics[width=\linewidth]{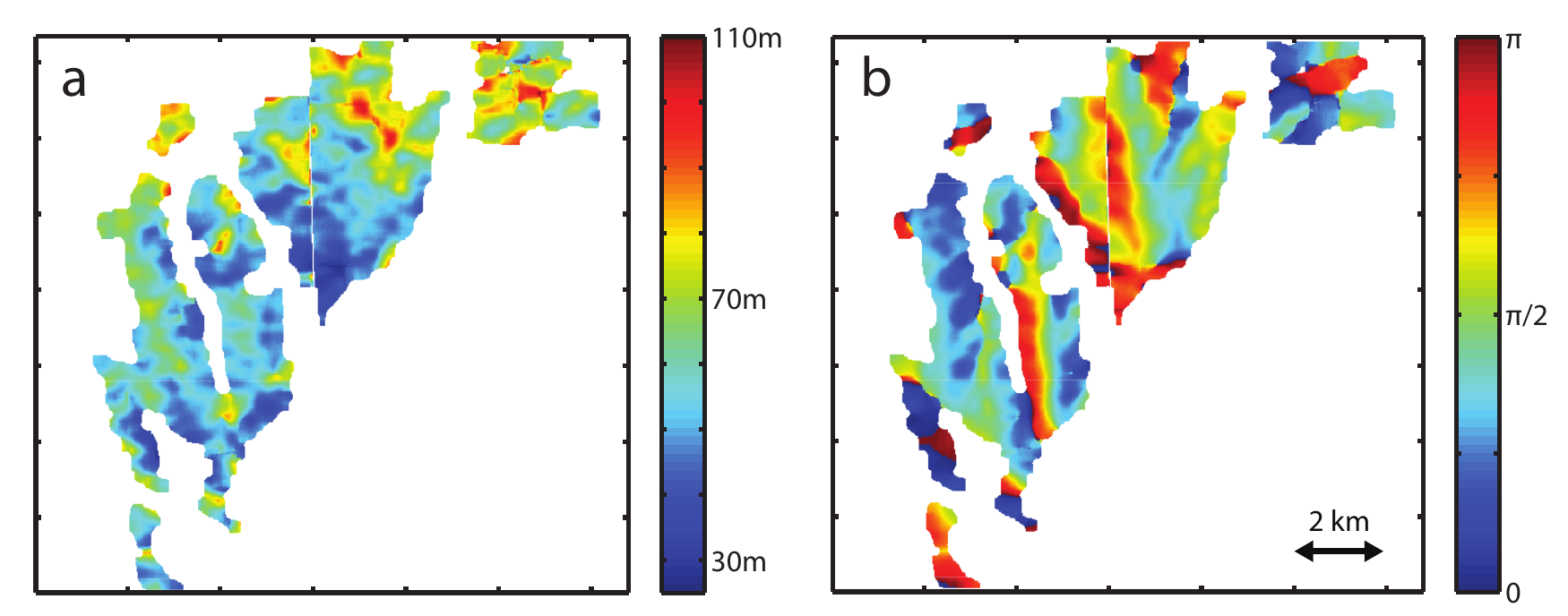}
\end{figure}

The local Fourier analysis was run twice, using windows with $L= 260$~m and
$L=400$~m, applying the window to overlapping regions on 20~m intervals, and
generating smoothly-varying maps of the local pattern properties.  The two
window sizes ensured that the largest wavelengths could be analyzed ($L=400$~m),
and exploited the greatest feasible localization given the median pattern
wavelengths ($L=260$~m).  The $L=400$m method identified large $\lambda$ in regions where the $L=260$m method identified  no pattern; while the $L=260$m method allowed the pattern properties to be identified in regions with dimensions $< 400$~m.  To combine the outcome, we averaged the results of the analyses for overlapping cells (resulting in a minimum change in the identified $\Theta$ and $\lambda$), and retained the uniquely identified $\Theta$ and $\lambda$ in cells where either mechanism failed. Since the pattern properties in windows located near the edges of the
pattern are influenced by both patterned and unpatterned regions, we trimmed
the resulting $\lambda(x,y)$ and $\Theta(x,y)$ fields by 130~m to discard the most
strongly-affected regions. The results of the complete analysis are shown in
Fig.~\ref{fig:fourier}.

We developed a map of the soil type on the 20~m grid by interpolating the
SSURGO data. We smoothed the NED elevation data to remove high-resolution
artifacts \citep{Oimoen-2000}, and computed the topographic gradient (steepness)
and aspect (orientation) over the same 20~m grid.  We made direct comparisons
between $\lambda(x,y)$, $\Theta(x,y)$ and the soil and topographic features. To
cope with the large, noisy dataset, we grouped the data into bins of equal size and related the median and interquartile range of the local
pattern characteristics to the predictor variables across the bins. The behavior
of these summary statistics and the dataset was analyzed using least squares regression.  
We analyzed the frequency of occurrence of deviations between the pattern and hillslope orientations $\Delta \Theta$ and the frequency of occurrence of non-unique pattern
orientation and wavelength $Q_\lambda$ and $Q_\Theta$ as a function of location
(e.g. near topographic minima (drainage lines), maxima (ridges), anthropogenic
features (roads and trails), and the edges of the pattern). 

\subsection{Comparison to models}
To quantify the degree to which local pattern variations can be explained as non-ideal pattern features that can arise from the nonlinear dynamics of the pattern forming mechanisms, we performed a control analysis using simulated data. Our
test region is a $500 \times 800$~m$^2$ region of the 0.5~m  resolution pattern.  Analyzed at the $200 \times 200$~m$^2$ scale, the pattern in this region contained a single high-energy wavenumber and direction peak, indicating that it  could be reasonably modeled with homogeneous model parameters.  We used the pattern in this region as the initial condition for a physically-based vegetation band-forming model \citep{Rietkerk-2002, Thompson-2009}.  We used this model because it (a) represents the surface runoff feedback mechanisms that were observed to occur at this site by \citet{McDonald-2008}; and (b) simulates vegetation bands that retain curvature, variability and other non-ideal behaviors (compared to more idealized models \citep{Lefever-1997}) and thus has the potential to generate spatially variable pattern properties endogenously.  We selected model parameters by stepping through reasonable parameter combinations and selecting the combination that minimized the variation between the observed vegetation pattern and the model prediction after 5000 timesteps \citep{Thompson-2008}.  This method allows us to identify a parameter set for the model that approximates the observed patterns as a stable steady state solution.  Using both the model output and the original binary images, we apply the same Fourier windowing method and compute key spatial statistics (mean, variance, and autocorrelation length, taken as the lag at which the autocorrelation halved) for the resulting $\lambda(x,y)$ and $\Theta(x,y)$.

\section{Results \label{sec:results}}  

\subsection{Local properties of vegetation patterns}

Using the Fourier windowing method, we find that 44\% of the 188 km$^2$
area contains patterned vegetation with a clearly identifiable wavelength and
orientation.  The Fourier windowing method fails to detect some areas
where vegetation
patterns are visually identifiable.  These include regions where vegetation
bands are confined to fingers of a particular soil type that are 
$\lesssim 260$~m in extent, meaning that the pattern cannot be identified
as the dominant Fourier mode in any given window.  There are also some regions
where a pattern can be identified by eye, but is so disordered that it falls
below the noise threshold.
Approximately 10\% of the study area consists of isolated patches of patterning
which offer little opportunity to explore spatial variations. Instead, we
confine  our analysis to the 34\% ($\approx 64$~km$^2$) of the image that consists of
spatially-connected vegetation patterns.  Within these regions, the pattern has
a mean wavelength of $\bar{\lambda} = 63\pm14$~m (reported variations are
standard deviation unless otherwise specified) and an autocorrelation length of
approximately 800~m (i.e. $\approx 12 \lambda$).  The probability distribution
function (PDF) of $\Theta(x,y)$ values is peaked in the north-south
direction, but spans a full $0\leq \Theta<\pi$ range.  Table~\ref{tab:Stats}
provides summary statistics describing the average pattern characteristics,
topography and soil properties.

While seven distinct soil types occur in the study area, two of
these, the Delnorte and Reakor associations, contain 97\% of the vegetation
patterning 
These soils are distinguished by a relatively
low clay and high silt content.  The Delnorte association is characterized by a very
shallow soil depth (23~cm) due to the presence of calcium-carbonate based
hardpan or petrocalcic horizon. By contrast, Reakor association soils are at
least 2~m deep. The full range of soil properties is shown in Table~\ref{tab:Soils}.

\begin{figure}[h]
\caption{ (a) Frequency distribution of $\Delta\Theta$, the deviation between
the local pattern orientation and the local topographic orientation.  The slope orientations deviate by less than $\pi/8$ for 80\% of the image.  (b) Comparison of the local pattern
wavelength $\lambda$ to the local topographic slope. Data were collected into
10 bins of equal size ($\approx 11000$ data points per bin). 
The box and whisker plots indicate
the median (central lines); the 25th and 75th quartiles (box limits) within the bin.
The whiskers extend two inter-quartile ranges from the median.  The thick red
line indicates a linear fit to the medians.  (c) PDF of 
the pattern wavelengths associated with the two major soil classes, indicating
the bimodality in $\lambda$ on Reakor association soils.} 
\label{fig:controls}
\includegraphics[width=\textwidth]{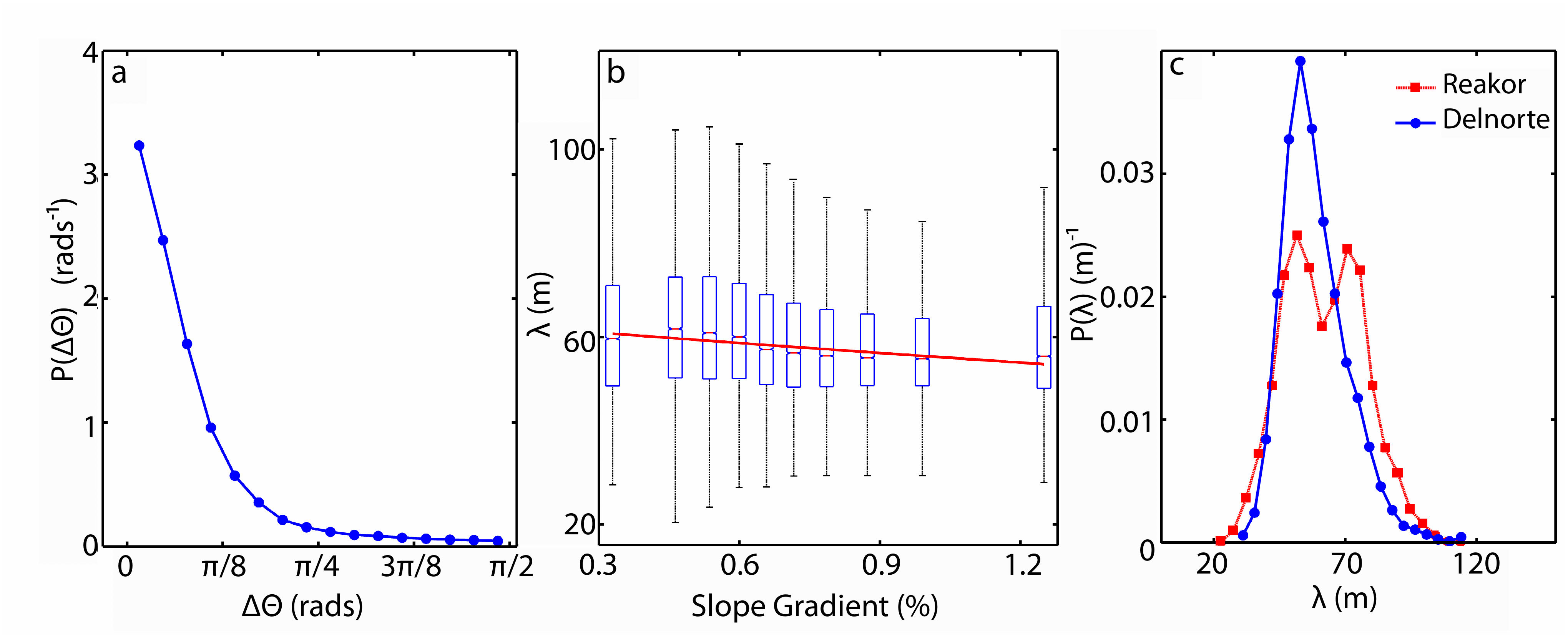}
\end{figure}

We find that three factors broadly determine the local pattern properties: the
orientation of the slope, the steepness of the slope, and the soil type. The
orientation $\Theta$ of the pattern is almost completely determined by the
underlying slope orientation. As illustrated in Fig.~\ref{fig:controls}a,
approximately 81\% of the pattern is oriented within $\pm \pi/8$ radians of the
hillslope. Deviations between the pattern and hillslope orientations, denoted
$\Delta\Theta$, are discussed further in \S\ref{sec:results}.\ref{sec:NUPA}.
Second, we observe a significant relationship between the local
pattern wavelength $\lambda$ and the hillslope gradient.
Fig.~\ref{fig:controls}b illustrates the moderately strong, significant
decline in the median wavelength $\lambda$ of equally sized data bins
 with increasing hillslope gradient ($r^2 = 0.63, p = 6 \times
10^{-3}$).  
In addition, we found that the pattern wavelength is influenced by the soil type.
Fig.~\ref{fig:controls}c shows PDFs of $\lambda$ for each soil type. While
Delnorte soils have a unimodal PDF (mode $\lambda = 53$~m), the Reakor
association soils have a bimodal distribution. One mode corresponds to that of
the Deltnorte soils ($\lambda =52$~m), but the other modal wavelength is much
longer, with $\lambda = 71$~m. Further analysis of this effect is provided in
\S\ref{sec:results}.\ref{sec:soil}

\subsection{Comparison to models}
The results are shown in  Table~\ref{tab:Stats2}. We find that while the mean properties of the model and observed pattern are similar (a consequence of calibrating  the model to these means), $\Theta$ and $\lambda$ vary $3-5$ times more in the observed pattern than in the modeled pattern.  The underlying topography is still more variable, and presumably causes the variability of the observed pattern. Thus, even in the region of the study site with the greatest uniformity in the pattern properties, the spatial heterogeneity observed in the real patterns exceeded the pattern variability that could be produced by a physical model.  These results justify our attribution of the remaining variability in the pattern properties across the site to environmental variation rather than to defects or initial condition effects arising from the nonlinear dynamics of the system.

\subsection{Non-uniqueness in pattern attributes \label{sec:NUPA}}

\begin{figure}[h]
\caption{Spatial distribution of the uniqueness metric  $Q_\Theta$ for the
pattern orientation,
and magnified cases illustrating different examples of non-uniqueness in the
pattern orientation.  Panel (a) indicates non-uniqueness in orientation associated
with a road, which adds north-south oriented structure to the local
NE-SW oriented banding due to upslope ponding of runoff.  Panel (b)
indicates a region of changing pattern orientation and wavelength associated with
the interleaving of two contrasting soil types.  Panel (c) illustrates the short length-scales on
which the pattern can change orientation as it bends around a sharp ridge
(location indicated in blue). A proportion of the region where $\Delta\Theta\approx \pi/4$, as indicated by the different director arrows, has $Q_\Theta>0.75$. Panel (d) indicates the
deviation between pattern and slope orientation around a stream channel (estimated
location shown in blue); again there are regions where $\Delta\Theta\approx
\pi/2$ and $Q_\Theta>0.75$.}
\label{fig:qualitymap}
\includegraphics[width=\textwidth]{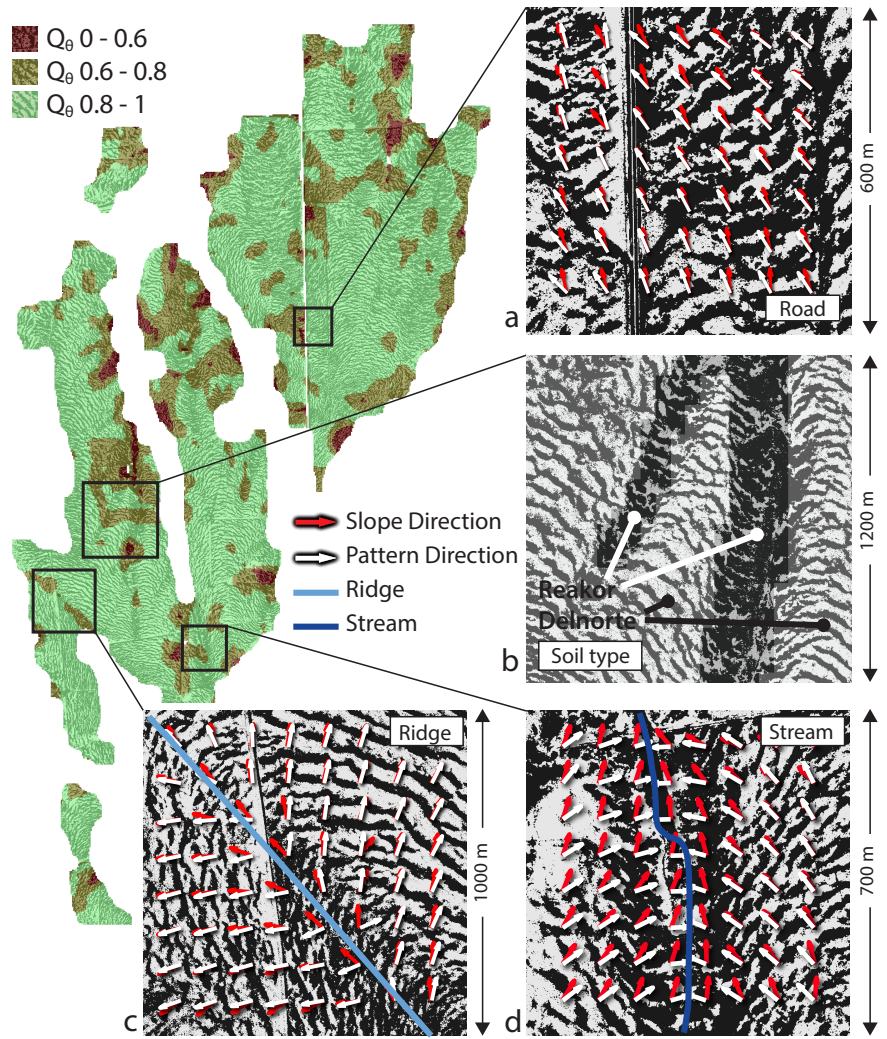}
\end{figure}

\begin{figure}[h]
\caption{Relative increase in frequency of (a,b) non-unique pattern
orientations and (c) wavelengths, for windows within
250~m of local elevation minima (streams), maxima (ridges), roads, and pattern
boundaries.  All frequencies are measured relative to the frequency in the
remainder of the pattern.
Non-unique orientations, for example, occurred 50-60\% more often near streamlines
and ridges than in the bulk of the pattern.}
\label{fig:qualitydist}
\includegraphics[width=\textwidth]{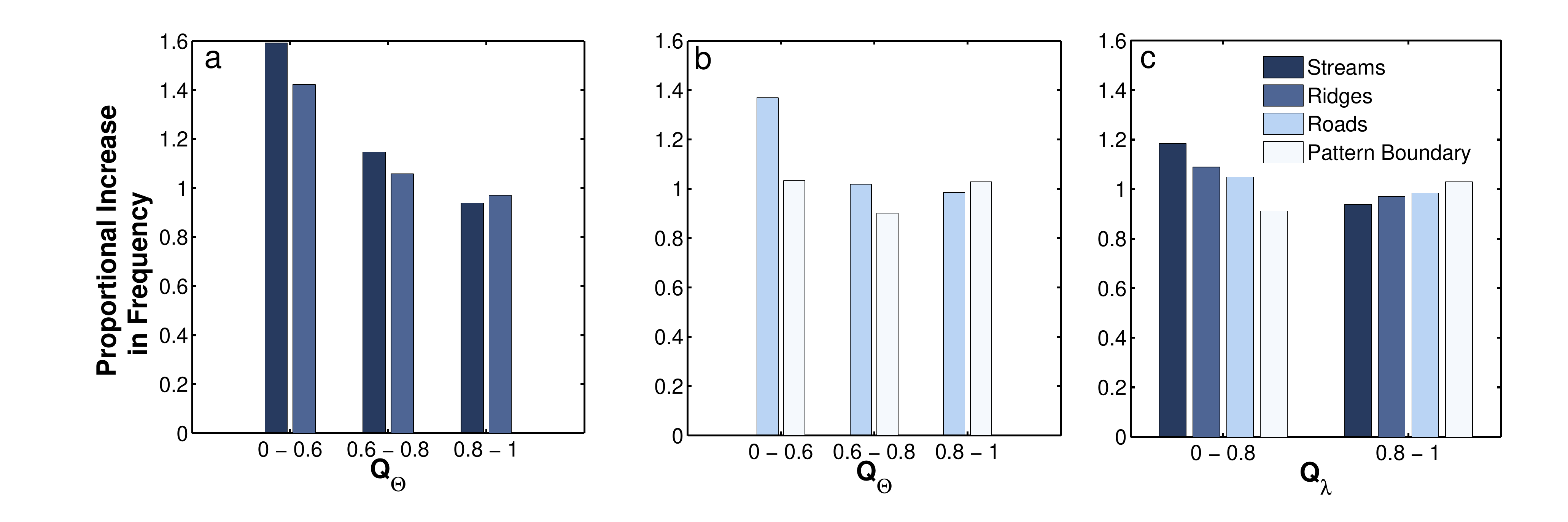}
\end{figure}

While the local pattern wavelength and orientation is on average set by the
local hillslope gradient and orientation, we nonetheless observe regions with
significant deviation from the overall trend. As shown in
Fig.~\ref{fig:controls}a, the deviation between pattern and hillslope orientation, $\Delta\Theta$, is $>\pi/8$ over approximately 20\% of the
pattern. There are even regions in which $\Delta\Theta \approx
\pi/2$, where the vegetated bands run parallel rather than perpendicular to the local hillslope
gradient.
Fig.~\ref{fig:qualitymap} shows the spatial distribution of the
orientation uniqueness metric  $Q_\theta$ across the patterned region,
highlighting regions of non-uniqueness. About half of the windows with large
$\Delta\Theta$ also contain more than one pattern direction (non-uniqueness in
$\Theta(x,y)$).  As illustrated by Fig.~\ref{fig:qualitydist}, non-unique
pattern orientations are clustered near streams (50\% increase in frequency of
$Q_\theta<0.75$ relative to the remainder of the pattern), ridges (50\% increase
in frequency), roads (30\% increase in frequency), and the pattern edge. We also observe regions with large deviations ($\Delta\Theta >
\pi/8$) but nonetheless unique pattern orientations ($Q_\theta \geq 0.75$). Such regions
also occur near ridges and streams more frequently than in the remainder of the
pattern ($\approx30\%$ more frequent in each case). 

The insets in Fig.~\ref{fig:qualitymap} illustrate regions of complexity in
$\Theta\left(x,y\right)$.  Fig.~\ref{fig:qualitymap}a and
Fig.~\ref{fig:qualitymap}b show regions of non-unique pattern directions.  In
Fig.~\ref{fig:qualitymap}a, we show a perturbation of the local pattern by a
road.  As illustrated, the upslope edge of roads is globally
associated with increased vegetation, while the downslope edges are typically bare.  Roads thus generate linear features that can create
non-uniqueness in the local pattern orientation. This panel also shows a second anthropogenic feature, a storm drainage outlet, which further decreases $Q_\Theta$ west of the road. 
Fig.~\ref{fig:qualitymap}b  illustrates how changes in soil type can
lead to low $Q_\Theta$. In this example, fingers of the Reakor association
soils are interleaved with the Delnorte
association soils, causing rapid changes in pattern wavelength and orientation,
and consequently multiple energetic values of $\Theta$ and $\lambda$ within the
260~m windows.  Fig.~\ref{fig:qualitymap}c illustrates a region where the
pattern orientation changes rapidly, turning
through approximately $\pi$ radians within a single 260~m window.  Such rapid
change
inevitably leads to low $Q_\Theta$ because there are multiple pattern
orientations located in a single window.  There are also, however,
regions near the ridge crest where $Q_\Theta\approx 1$ and the pattern is locally oriented
perpendicular to the slope. Fig.~\ref{fig:qualitymap}d is representative of
the complex vegetation transitions that occur near stream channels.  Here, too,
the pattern lies perpendicular to the local hillslope orientation, and rather
than curving into the streamline, remains broadly aligned with the band
orientation away from the stream.

\subsection{Soil type effects \label{sec:soil}}

\begin{figure}[h]
\caption{Distribution of the soils within the pattern
forming areas with respect to landscape position, referenced to the
nearest stream channel or local elevation minimum for (a) Reakor and (b) Delnorte Association soils.  Although both soil types
occur across the range of landscape positions, the Reakor association soils are
more strongly associated with riparian areas.  Distribution of pattern wavelengths on the Reakor Association (c)  and Delnorte Association (d) soils with respect to the nearest stream, again based on equally-sized
bins of the dataset.  There is a strong ($r^2 = 0.96$) and significant
($p<5\times10^{-7}$) decline in the wavelength on Reakor soils when moving from
the riparian areas to the uplands. There is no correlation between
landscape position and pattern properties on the Delnorte association
($r^2=0.08$ and $p=0.4$).}
\label{fig:StreamSoilWL}
\includegraphics[width=\textwidth]{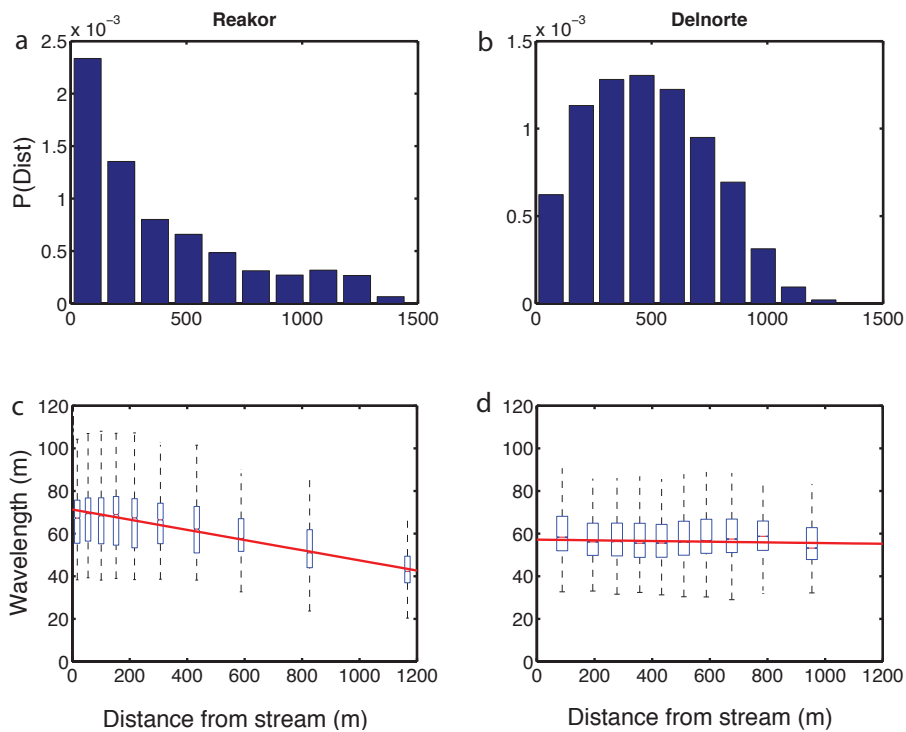}
\end{figure}

While the vegetation patterns on the Delnorte soils have a unimodal
wavelength distribution, the Reakor soils exhibit a bimodal distribution (see
Fig.~\ref{fig:controls}). The pattern associated with the second peak of
$\lambda$ on the Reakor soil consists of bands of bare soil within a matrix
of vegetation cover, inverting the distribution in the remainder of the pattern.
Visually, this `anti-stripe'
pattern is more disordered than the remainder of the pattern seen on the Reakor
association soils.  In the power spectra, these anti-stripe peaks contain less
than half the power of the short wavelength, coherent peaks.  Thus, on the
Reakor association soils the pattern varies through space from short to long
wavelength, ordered to disordered patterns, and lower to higher biomass.
Because increased biomass in drylands implies an increased access to water, we
examine how $\lambda$ varies as a function of the distance to the nearest stream
(as a proxy for availability of water). The results are shown in
Fig.~\ref{fig:StreamSoilWL}. While patterns on the Delnorte soils did not show
a distance-dependent $\lambda$ ($r^2=0.08$ and $p=0.4$), the patterns on  Reakor
soils show a trend of increasing $\lambda$ with decreasing distance to the
nearest stream ($r^2 = 0.96$ and $p<5\times10^{-7}$).

\section{Discussion \label{sec:discn}}  

The local Fourier metrics confirm Hypothesis (1) by showing that the wavelength
and orientation of the vegetation patterns are typically  coherent on
scales of $600-800$~m, but can change on scales of $\approx 20$~m when the
slope orientation changes over the same scale. We were unable to reproduce observed variation with a model even within regions of relatively uniform patterning, confirming Hypothesis (2) and further suggesting that exogenous factors rather than the pattern forming mechanisms drive the
spatial variability in $\lambda$ and $\Theta$. In light of these findings, and to address Hypothesis (3),
our subsequent analysis focuses on the relationship between exogenous factors
and the local pattern properties.

Our analysis indicates a hierarchy of controls on the morphology of the
vegetation patterning: at a global scale the pattern morphology is determined
by the slope orientation and the hillslope gradient, while the soil type imposes the
template for the pattern forming region.  At smaller scales, complexity in the form of multiple
local pattern orientations and deviations from the global trends are associated
with ridges, streams, roads and changes in the soil type.  Organization of the
pattern characteristics on small scales arises due to a combination of soil type
and topographic context: homogeneous, well defined patterns with a unimodal
$\lambda$ on Delnorte soils, regardless of the topographic context.  On Reakor
association soils, the pattern $\lambda$ increases in locations closer to
streams, while the pattern simultaneously becomes more disordered and vegetation
cover increases.

Several of these relationships support previous observations. The average
hillslope gradient of 0.7\% at the Fort Stockton site conforms to observations and
predictions of a minimum hillslope gradient being
needed for band formation \citep{Ursino-2006}, and exceeds the minimum slope
gradient observed in other settings (e.g. 0.25\% in the Sudan
\citep{Deblauwe-2011}). In addition, the inverse 
relationship between pattern wavelength and the hillslope gradient (see
Fig.~\ref{fig:controls}) has been identified elsewhere \citep{Valentin-1999,
dHerbes-1997, Eddy-1999, Deblauwe-2011}. However, the one model
\citep{Klausmeier-1999} that has been used to study wavelength-gradient
correlations predicts the {\itshape opposite} trend:
steepness causes larger wavelengths
\citep{Sherrat-2005, Ursino-2006}. We explored this trend with a different model \citep{Rietkerk-2002}  and find that it also produces
the opposite trend to that observed in natural systems.
While both models predict that multiple pattern wavelengths
are stable for a given hillslope gradient
\citep{SherrattLord-2007, Thompson-2009}, the selection of particular longer or
shorter wavelengths within that range is clearly at odds with observations.

The close correspondence between hillslope and pattern orientation at the Fort
Stockton site is a near-universal feature of vegetation patterning. The
observed deviations generally arise due to local effects that disrupt the
pattern, rapidly alter the direction of water flow, or change the
strength and/or length-scales of the pattern forming feedbacks.

Pattern disruption is exemplified by the effects of roads
(see Fig.~\ref{fig:qualitymap}a).
Roads disconnect upslope and
downslope vegetation bands by preventing the redistribution of water.  Rapid
alteration of the direction of water flow
occurs along ridges and stream
channels.  The pattern near these locations is less likely to have a unique
orientation or wavelength than in the mid-slope areas, and is more likely to have a
large $\Delta\Theta$.  The prominent ridgeline shown in
Fig.~\ref{fig:qualitymap}c is locally surrounded by a region with
$\Delta\Theta\approx\pi/2$ and $Q_\theta>0.75$.  This location provides an
unambiguous example of a change in pattern orientation near no-flow boundary
conditions, as predicted by \citet{McGrath-2012}.  Other ridgelines are
associated with large $\Delta\Theta$, but the low $Q_\Theta$ in these locations
makes the interpretation of the observed $\Delta\Theta$ ambiguous.  Methods that
identify a truly local metric of pattern properties, instead of the quasi-local metric
used here could both help to resolve these ambiguities, and to extend the
analysis into patterned areas less than 260~m wide.

Stream channel locations are also associated with rapid, and sometimes
discontinuous changes in pattern orientation. Like the ridge shown in
Fig.~\ref{fig:qualitymap}c, the stream channel shown in
Fig.~\ref{fig:qualitymap}d is surrounded by a region where
$\Delta\Theta\approx\pi/2$ and $Q_\theta>0.75$, i.e. a unique pattern
orientation that is perpendicular to the one expected from the slope
orientation. There has been little
exploration of the interaction of vegetation pattern with stream channels,
presumably because the current paradigm of vegetation pattern models offer
little reason to think that such interactions would be important.  Stream
channels occur in locations where surface runoff rapidly flows away and therefore cannot affect  vegetation upslope from
the channel. However, we  find evidence that the distance from a stream
channel alters the pattern properties on the Reakor Association soil type.  On
these soils the pattern wavelength, vegetation cover and disorder increase near
the streams.  These observations suggest that an additional mechanism could
affect the patterning at this study site: for instance, the stream-channel
boundary condition propagating back up into the hillslope.

We hypothesize that the ultimate cause of the changes in pattern morphology on
the Reakor soils lies in the contrast in the soil depth between the
Delnorte and Reakor associations: from 23~cm to over 2~m.  We are not the
first to propose an association between soil depth and the vegetation pattern
structure.  Depth to a silcrete hardpan is associated with a transition between
sharp patterns (shallow hardpan) and diffuse patterns (deep or no hardpan) in
Australia \citep{Mabbutt-1987, Tongway-1990}.  Strikingly coherent vegetation
bands in Niger occur above a shallow ironstone hardpan \citep{White-1970}.  The
broad, diffuse banding near Fort Stockton studied by \citet{McDonald-2008}
occurs on deep clays. \citet{McDonald-2008} cited unpublished research claiming
that vegetated bands were associated with a local increase in the depth to the
hardpan, similar to previous observations of increased vegetation density on
deep soils in Australia \citep{Mott-1974}.

\begin{figure}[h]
\caption{Schematic illustration of possible
effects of a shallow impeding layer on pattern formation mechanisms.
(a) In the
absence of an impeding layer, plants develop deep root systems to exploit water
from across the whole soil profile
(b) A shallow impeding layer
restricts the vertical growth of roots, confining them to the surface soils, and
promoting root competition laterally.
(c) The shallow impeding layer also
creates the potential for complex subsurface hydrology.  Saturation of soils
above the impeding layer may promote saturation excess runoff.  Gradients - due
to changes in the depth of the impeding layer, or in water potential across the
impeding layer - can also induce saturated porous media flow.  Provided that
surface soils near plants can freely drain, the altered hydrology may exaggerate
the positive feedback between plants and water.}
\label{fig:Conceptual}
\includegraphics[width=\textwidth]{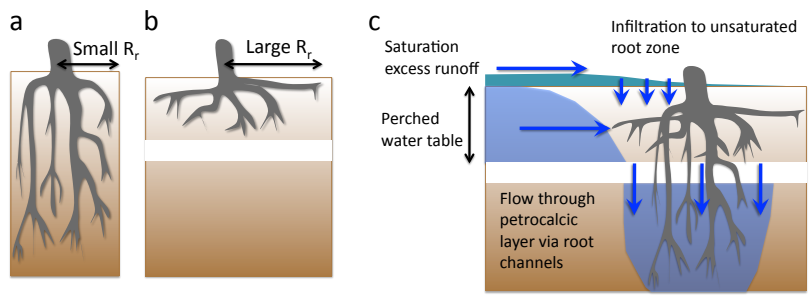}
\end{figure}

Field evidence is needed to determine the exact mechanisms by which soil depth
and proximity to  streams could lead to the changes we observed in pattern
morphology. Three scenarios illustrating potential mechanisms are shown in
Fig.~\ref{fig:Conceptual} and could form the basis for future field studies.
First, shallow soils could promote lateral root extension by plants,
exaggerating the effects of root competition \citep{Gilad-2004, Yizhaq-2005}
(see Fig.~\ref{fig:Conceptual}ab).  Studies of Chihuahuan desert species
confirm that there is intense root competition in the zone above the petrocalcic
horizon, where root growth is concentrated \citep{Gibbens-2001}.

In the second scenario, we recognize that shallow surface soils have limited water storage and saturate readily. For example, the Delnorte Association has as little as 2~cm of water storage capacity in the soils above the impeding layer.  
Unlike dry soils, which only generate runoff during
very heavy rains, saturated soils shed all rainfall as runoff. If soils did not
saturate near plants this runoff water could be trapped and infiltrate in these
locations (see Fig.~\ref{fig:Conceptual}c). Plant roots penetrate and thus may
break up hardpans \citep{Gibbens-2001}.  Root water uptake is also a driver of
hardpan formation \citep{Duniway-2007}, and hardpans might thus form at
greater depth beneath deep-rooted (woody) vegetation.  Either mechanism could
prevent the surface soil from saturating near the vegetated bands and allow it to store runoff.

The third scenario also relates to the
potential for saturation to occur above the hardpan, allowing subsurface saturated flow to occur (see
Fig.~\ref{fig:Conceptual}c). The changes in pattern properties with distance to
the stream suggests that water availability increases downslope: this
requires subsurface storage, if not flow.  A relationship between the pattern
formation and subsurface flow could explain the local deviations of pattern
orientation from slope orientation near the streamlines since the surface
orientation might not correspond to the local water table flow direction.

All three scenarios would tend to strengthen the pattern forming feedbacks on
the shallow soils. They could be investigated using subsurface soil moisture
sensors to observe shallow soil saturation; water isotope tracers to determine
the water sources used by
plants, observations of calcium ion concentrations in runoff which would provide
an indicator of water contact with the petrocalcic horizon and root excavations to
compare morphologies in sites with different depths to the impeding layer.
These studies could be valuable to provide more information about the
hydrological role of petrocalcic horizons, which have received relatively little
attention given their ubiquity in desert environments \citep{Duniway-2007}.

\section{Conclusion \label{sec:conc}}  

Large scale analyses of variation in pattern morphology has provided broad
confirmation of many theoretical predictions about vegetation patterning and its
variation along climatic gradients \citep{Deblauwe-2011}, which are consistent
with our observations at local scales.  Our analysis shows that while
pattern properties mostly vary on scales of $600-800$~m, they can change much
more rapidly around boundaries in topography or soil characteristics.  By
analyzing the pattern on these fine scales, we identified
deviations between hillslope and pattern orientation, soil-controlled changes in
pattern wavelength, coherence and vegetation cover, and at least one likely
example of the boundary condition effects predicted by \citet{McGrath-2012}.
Two observations, echoed at multiple other sites, are not well-explained by
current theory: the observation of increasing pattern wavelength on steepness,
and the soil controls on vegetation pattern length-scale and coherence.

The large timescale separation between individual storm events and the timescales
on which desert vegetation distributions change means that an ongoing
dialogue between empirical and theoretical studies is critical for understanding
the dynamics of these ecosystems.  Local information about pattern
qualities, when combined with high resolution information about the pattern
substrate, is evidently a useful additional tool for analysis.

Despite the advances in understanding vegetation patterns in the past ten years,
theoretical models still require the use of effective parameters to describe
feedbacks, soil-plant-water interactions and the resulting landscape fluxes.
Linking theory and observation to make quantitative predictions, therefore, remains an outstanding challenge.  Addressing this challenge requires improved
observations of within-storm hydrologic processes and plant water use:
observations that ecohydrologists are increasingly equipped to make due to
developments in distributed sensing systems and tracer technologies.
Computational tools for assessing three-dimensional soil moisture dynamics,
land-atmosphere interactions and vegetation spread are also improving. By
coupling these tools with detailed field measurements, there is potential to develop a detailed theoretical
framework that can address the consequences of changing soil depth, root
orientation, runoff generation mechanisms and subsurface flow processes on the
overall dynamics and resilience of
vegetation patterns.

\section{Acknowledgements}
SET and GP acknowledge support from NSF-EAR-1013339.

\renewcommand{\baselinestretch}{1}


\newpage
\section{Tables}

\begin{table}[h]
\caption{Pattern characteristics of natural patterning across
the full pattern extent.}
\label{tab:Stats}
\begin{tabular}{|c|c|c|c|}
\hline ~ &Mean&Std. Dev.&Corr. Length (m) \\
\hline Pattern wavelength, $\bar\lambda$ (m)&63&13.8&800 \\
Mean pattern orientation, $\bar\Theta$ (rads)&1.4&0.76&600 \\
Hillslope orientation (rad)&1.2&0.75&1400 \\
Hillslope gradient &0.007&0.002&1900 \\
\hline
\end{tabular}
\end{table}

\begin{table}[h]
\caption{Soil properties for the Delnorte and Reakor Association Soils.  Note that the soil textural percentages do not sum to 100 because gravel and organic content have not been reported.}
\label{tab:Soils}
\begin{tabular}{|c|c|c|}
\hline & Reakor & Delnorte \\ \hline
Soil depth (cm)		&   200	& 23	\\
Hydraulic conductivity ($K_{sat}$, $\mu$m~s$^{-1}$)	& 2.1	&59.2	\\
Clay content (\%)			& 31.5	& 8.7	\\
Sand content	  (\%)		& 6.8 	&	42.8\\
Silt content	 (\%)		& 61.7	&	21.1\\
\hline
\end{tabular}
\end{table}

\begin{table}[h]
\caption{Pattern characteristics comparing analysis of natural and simulated
data for a a $500 \times 800$~m$^2$  region.
The simulated data comes from a  modified Rietkirk model calibrated to match
the mean pattern properties of the natural data.}
\label{tab:Stats2}
\begin{tabular}{|c|c|c|c|}

\hline ~ &Mean&Std. Dev.&Corr. Length (m) \\\hline
Natural pattern wavelength $\bar\lambda$ (m)&53&4.1&350 \\
Model pattern wavelength $\bar\lambda$ (m)&54&1.7&360 \\
Natural pattern orientation  $\bar\Theta$ (rad)&1.7&0.39&420 \\
Model pattern orientation $\bar\Theta$ (rad)&1.6&0.07&390 \\
Hillslope orientation (rad) &1.7&0.5&120 \\ \hline
\end{tabular}
\end{table}

\section{Figures}

h

\end{document}